\documentclass[prb,twocolumn,showpacs,preprintnumbers,amsmath,amssymb,superscriptaddress]{revtex4}
\usepackage{amssymb}
\usepackage{color}

\def \Grenoble{Laboratoire National des Champs Magn\'etiques Intenses (LNCMI), CNRS-UJF-UPS-INSA, 38042 Grenoble, France}
\def \Prague{Institute of Physics, Faculty of Mathematics and Physics, 12116 Prague, Czech Republic}
\def \Regensburg{University of Regensburg, Regensburg, Germany}
\def \Warsaw{Institute of Physics, Polish Academy of Sciences, 02-668 Warsaw, Poland}

\usepackage{epsfig}
\usepackage{amsmath}
\begin{document}
\title{Magneto-resistance quantum oscillations in a magnetic two-dimensional electron gas}
\author{J. \surname{Kunc}}
\email{kunc@karlov.mff.cuni.cz}
\affiliation{\Grenoble}
\affiliation{\Prague}
\author{B. A. \surname{Piot}} \affiliation{\Grenoble}
\author{D. K. \surname{Maude}} \affiliation{\Grenoble}
\author{M. \surname{Potemski}} \affiliation{\Grenoble}

\author{R. \surname{Grill}} \affiliation{\Prague}
\author{C. \surname{Betthausen}} \affiliation{\Regensburg}
\author{D. \surname{Weiss}} \affiliation{\Regensburg}
\author{V. \surname{Kolkovsky}} \affiliation{\Warsaw}
\author{G. \surname{Karczewski}} \affiliation{\Warsaw}
\author{T. \surname{Wojtowicz}} \affiliation{\Warsaw}

\date{\today}

\begin{abstract}
Magneto-transport measurements of Shubnikov-de Haas (SdH)
oscillations have been performed on two-dimensional electron gases
(2DEGs) confined in CdTe and CdMnTe quantum wells. The quantum
oscillations in CdMnTe, where the 2DEG interacts with magnetic Mn
ions, can be described by incorporating the electron-Mn exchange
interaction into the traditional Lifshitz-Kosevich formalism. The
modified spin splitting leads to characteristic beating pattern in
the SdH oscillations, the study of which indicates the formation
of Mn clusters resulting in direct anti-ferromagnetic Mn-Mn
interaction. The Landau level broadening in this system shows a
peculiar decrease with increasing temperature, which could be
related to statistical fluctuations of the Mn concentration.
\end{abstract}


\maketitle
\section{Introduction}

At low temperatures, the longitudinal resistivity of metallic
systems exhibits quantum oscillations when submitted to a
sufficiently high magnetic field. These so-called Shubnikov-de
Haas (SdH) oscillations  are, in particular, characteristic of
two-dimensional electron gases (2DEG) confined in semiconducting
structures and appear at low magnetic fields prior to the
development of the quantum Hall effect in clean systems. The
analysis of the magnetic field and temperature dependence of the
SdH oscillations
\cite{ArgyresPhysRev104-1956,LifshitsJPhysChemSolids4-1958,Ando74b,Isihara86}
provides a valuable information on the quantized density of states
(e.g. the Landau level shape,\cite{Ando74b,EndoJPSJ77-2008} and
the cyclotron and spin gaps \cite{PiotPRB72-2005} as well as on
the nature of the carrier scattering \cite{Isihara86,Ando74b} and
the associated quantum life times
.\cite{ColeridgePRB49-1994,ColeridgeSS361-1996} While the most
detailed studies were historically undertaken in high mobility
GaAs-based 2DEG, other 2D systems of high quality have slowly
emerged, enabling us to explore the influence of different
parameters such as the valley
 \cite{Klitzing1980,Takashina2006,Novoselov2004} and spin degrees
of freedom, \cite{Piot10,QHEZnO2007} as well as the effect of
magnetism in these systems.
\cite{SchollAPL62-1993,TeranPRL88-2002,BuhmannAPL86-2005}

In this work, we present an investigation of the SdH oscillations
in a high quality ``magnetic 2DEG'' in a diluted magnetic
semicondutor, CdMnTe. \cite{SchollAPL62-1993,TeranPRL88-2002}
CdMnTe is grown by substituting a small fraction of Cd atoms by Mn
in the original (non-magnetic) CdTe II-VI semiconductor. The high
quality of these systems was recently demonstrated by the
observation of the fractional quantum Hall effect.
\cite{Piot10,CdMnTeFQH2014} A systematic comparison of the SdH
oscillations in both systems is made here to identify the
particular effects related to the presence of magnetic Mn ions.
The SdH oscillations in CdTe exhibits a behavior similar to the
widely studied GaAs-based 2DEG; a field/temperature independent
Landau level broadening characteristic of long range scattering
mechanism, and an exchange-enhanced spin gap leading to spin-split
oscillations (a doubling of the frequency) above a critical
magnetic field. In CdMnTe, the SdH oscillation exhibit an
additional beating pattern with nodes where the oscillations have
a vanishing amplitude. We show that this behavior is a consequence
of the giant Giant Zeeman splitting (GZS) resulting from the $s-d$
exchange interaction between electrons and the $S=5/2$ Mn spins.
\cite{TeranPRL88-2002} The SdH characteristics can be
well-described by incorporating the electron-Mn exchange
interaction into the traditional Lifshitz-Kosevich formalism.
\cite{LK56} For a good quantitative description, the formation of
Mn pair clusters with direct Mn-Mn antiferromagnetic interactions,
which reduce the average Mn spin polarization, has to be
considered. Another peculiarity of the magnetic 2DEG is a decrease
of the Landau Level broadening with increasing temperature,
together with an increase in the broadening with increasing
magnetic  field. This suggests a connection between the Landau
level broadening and the Mn spin polarization, as expected in the
presence of local fluctuation in the Mn concentration.

\section{Samples}
The non-magnetic CdTe sample consist of a $20$~nm-wide CdTe
quantum well (QW), modulation-doped with iodine on one side, and
embedded between Cd$_{0.74}$Mg$_{0.26}$Te barriers. The magnetic
sample consist of a 21.1 nm-wide Cd$_{1-x}$Mn$_{x}$Te QW. The
average Mn concentration of $\sim0.3\%$ is introduced by
delta-doping within 7 separate monolayers among the 65 CdTe
monolayers composing the QW. The samples, in form of
1.5$\times$6mm rectangles, were fitted with electrical contacts in
a Hall bar-like configuration. Experiments have been carried out
in a $^3$He/$^4$He dilution refrigerator inserted into a
superconducting magnet. A standard, low frequency ($\approx
10$~Hz) lock-in technique has been applied for the resistance
measurements. The samples were illuminated by using the
514~nm-line of a Ar$^+$ laser to increase the 2DEG mobility. The
laser illumination was limited to $\sim50$~$\mu$W/cm$^2$ but
permanently maintained as it was found to assure the most stable
conditions over the different experimental runs (the resulting
heating effects on the 2DEG were estimated directly from the
magneto-resistance). A special attention has been paid to use slow
sweeps not to affect the amplitude of fast SdH oscillations. Under
our experimental conditions, the CdTe 2DEG density was $4.5\times
10^{11}$~cm$^{-2}$ (corresponding to a Fermi energy of
$E_F=10.8$~meV), with a low temperature mobility of $\mu=2.6
\times 10^{5}$~cm$^2$/Vs. The CdMnTe 2DEG density was $4.0\times
10^{11}$~cm$^{-2}$ (corresponding to a Fermi energy of
$E_F=9.6$~meV), with a low temperature mobility of $\mu=1.2 \times
10^{5}$~cm$^2$/Vs. The effective mass and the $g$-factor of
electrons in CdTe, $m_e=0.1m_0$ and $|g_e|=1.6$,  were determined
by far infrared magneto-absorption and Raman scattering
spectroscopy.

\section{Experimental results}\label{expRes}

\begin{figure}[h]
\centering
\includegraphics[width=9cm]{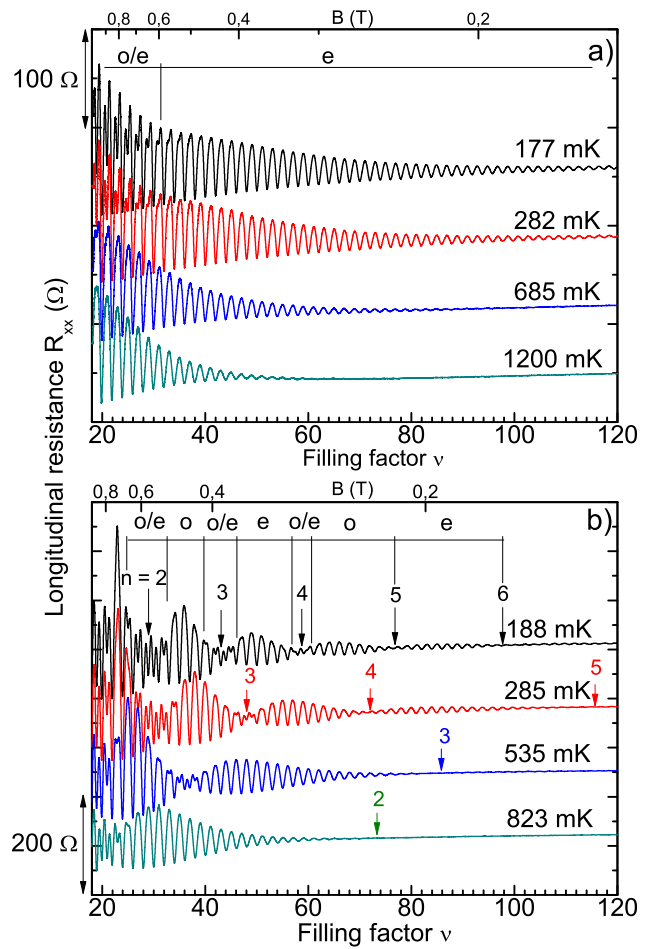}
\caption{Longitudinal magneto-resistance in (a) CdTe and (b)
CdMnTe QW for four selected temperatures. The data are shifted
along the y-axis for clarity and the units of (a) $100~\Omega$ and
(b) $200~\Omega$ are marked along the left side by a double arrow.
Minima corresponding to odd, even and alternating odd and even
filling factors are labeled ``o'', ``e'' and ``o/e'',
respectively. The nodes (condition
$\Delta_s=(n+1/2)\hbar\omega_c$) in the beating pattern of
magneto-resistance of CdMnTe QW are labeled by integer index $n$.}
\label{Fig1}
\end{figure}

The magneto-resistance of the 2DEG in CdTe and CdMnTe QWs is shown
for four selected temperatures in Fig.~\ref{Fig1}~(a) and (b),
respectively.
For the sake of comparison we plot the data as a function of the
filling factor $\nu=B_1/B$, where $B_1$ is the magnetic field at
filling factor $\nu=1$ ($B_1=18.8$~T in the CdTe and $B_1=16.5$~T
in the CdMnTe QWs). The magneto-resistance in the CdTe QW exhibits
the well-known SdH oscillations, which amplitude increases
(decreases) with magnetic field (temperature). At low magnetic
fields, before spin-splitting is observed the minima of the
longitudinal resistance $R_{xx}$ correspond to the situations
where the Fermi energy lies between two Landau levels (LL), in a
minimum of the total density of states $G_{tot}$. When the Fermi
energy lies in the center of a Landau level, maxima in $R_{xx}$
are observed. Above a critical magnetic field, electron-electron
exchange interactions lift the Landau level spin degeneracy,
\cite{PiotPRB72-2005} which leads to alternating odd and even
filling factors minima in the SdH oscillations (visible e.g. for
$\nu<30$ at $T=177$~mK in Fig.~\ref{Fig1}~(a)).

The magneto-resistance in the CdMnTe QW also exhibits SdH
oscillations, as can be seen in Fig.~\ref{Fig1}~(b). However, an
additional beating pattern is observed and ``nodes'' can be
distinguished in the SdH amplitude, as previously observed in
Ref.~\onlinecite{TeranPRL88-2002}. At low magnetic fields, the SdH
amplitude tends to zero in the region of the nodes, while at
higher fields, they are characterized by a local minimum of the
SdH amplitude associated with a doubled SdH oscillation frequency.
The presence of a strong electron-manganese exchange interaction
gives rise to a Giant Zeeman splitting (GZS) in the 2DEG, which
grows quickly as the localized Mn spins are polarized by the
applied magnetic field. The GZS saturates when the Mn spin
polarization has reached its maximum value, for a magnetic field
of typically $\sim 0.5$~T at low temperatures. This
``Brillouin-like'' strong field dependence of the GZS, compared to
the smaller linear increase of the cyclotron gap
($\hbar\omega_c=1.16$~meV/T), leads to the rather unusual
situation where the spin gap $\Delta_s$ can be several times
larger than the cyclotron gap. As the magnetic field increases,
the conditions $\Delta_s=n\hbar\omega_c$, where $n$ is a
(decreasing) integer, are successively satisfied. These magnetic
field-dependent commensurability of the spin and cyclotron gaps
leads to a maximized density of states when the Fermi level lies
in the center of coinciding (degenerate) levels. When the
spin-resolved Landau levels are all equally-spaced
($\Delta_s=(n+1/2)\hbar\omega_c$), the maximum density of states
is a factor of two smaller. The observed SdH beating are therefore
a direct manifestation of the GZS in magnetic 2DEGs. We note that
the ``nodes'' conditions($\Delta_s=(n+1/2)\hbar\omega_c$) should
also be accompanied by a doubling of the SdH frequency, similar to
the one observed at high enough magnetic fields in CdTe
(Fig.~\ref{Fig1}~(a)). This is indeed observed for small $n$ as
indicated by the down arrows in Fig.~\ref{Fig1}~(b). In the lower
magnetic field regime, where $\hbar\omega_c\leq\Gamma$, the
density of state modulation is so small that the nodes conditions
result in a disappearance of the SdH.

\section{Theoretical model}

In order to quantitatively describe the data in both CdTe and
CdMnTe, we have derived the formula describing the SdH oscillation
in the case of an arbitrarily large spin splitting $\Delta_s$. We
have used the Kubo-Greenwood
expression~\cite{Hill76,ColeridgeSS361-1996}
\begin{equation}
\sigma(B)=\int_{-\infty}^{\infty}\sigma(E)\left(-\frac{\partial n_{FD}}{\partial E}\right)dE
\label{equKuboGreen}
\end{equation}
to calculate the conductivity $\sigma(B)$ of electrons, where
$n_{FD}$ is the Fermi-Dirac distribution. The conductivity
$\sigma(E)$ calculated within the Drude model yields
$\sigma(E)=e^2n_{eff}/m_e\tau_{tr}\omega_c^2$ in the diffusion
limit ($\omega_c\tau_{tr}\gg 1$), where $n_{eff}$ is an effective
carrier concentration contributing to $\sigma(E)$, $m_e$ the
effective mass of electrons, $\tau_{tr}$ the transport life-time
and $\omega_c$ the cyclotron frequency.

The effective carrier concentration $n_{eff}$ is proportional to
the density of states at Fermi level $G(E_{F})$  and can be
written as $n_{eff}\propto\frac{G(E_{F})}{G_0}=1+\frac{\delta
G(E_{F})}{G_0}$, where $G_0=\frac{m_e}{2\pi\hbar^2}$ is the
zero-field density of states and $\delta G(E)$ is the modulation
of $G(E)$ for $B>0$~T ($G(E)=G_0+\delta G(E)$ and $\delta G(E)\ll
G_0$).

The relative change of the conductivity can then be written as
$\left|\frac{\sigma_{xx}(B)-\sigma_0}{\sigma_0}\right|=2^p\left|\frac{\delta
G}{G_0}\right|$ where the exponent $p$ depends on the type of
scattering ($p=1$ for long-range scattering and $p=2$ for short
range scattering where $1/\tau_{tr}\propto\frac{G(E_{F})}{G_0}$.
\cite{ColeridgegPRB39-1989}) $G(E)$ has been modeled as a sum of
either Lorentzian (Eq.\ref{lor}) or Gaussian (Eq.\ref{gau}) Landau
levels:

\begin{equation}
G(E)=\frac{m_e}{2\pi\hbar^2}\frac{\hbar\omega_c}{\pi\Gamma}\sum_{n=0}^{\infty}\sum_{s=\pm
1/2}\frac{1}{1+\left(\frac{E-E_{n,s}}{\Gamma}\right)^2}\label{lor}
\end{equation}

\begin{equation}
G(E)=\frac{m_e}{2\pi\hbar^2}\frac{\hbar\omega_c}{\sqrt{2\pi}\Gamma}\sum_{n=0}^{\infty}\sum_{s=\pm
1/2}\exp\left[{-\frac{(E-E_{n,s})^2}{2\Gamma^2}}\right],
\label{gau}
\end{equation}
where $\Gamma$ is the Half-Width at half Maximum (HWHM) of the Landau levels ($\Gamma=\hbar/2\tau_q$, where $\tau_q$ is the quantum lifetime) and $E_{n,s}$ is the energy of the LL with orbital (spin) quantum number $n$ ($s=\pm 1/2$). In order to compare model directly with the
resistance data, we have used the relation
$\left|\frac{\sigma_{xx}(B)-\sigma_0}{\sigma_0}\right|=\left|\frac{R_{xx}(B)-R_0}{R_0}\right|$
where $\sigma_0$ and $R_0$ are the zero-field conductivity and the
resistance, respectively, valid for a 2DEG in a quantizing
magnetic field. Hence, the final expression for the resistance
reads in the form of the Fourier series as:
\begin{eqnarray}
    \lefteqn{\left|\frac{R_{xx}(B)-R_0}{R_0}\right|=} \nonumber\\
    &&2^{p}\sum_{s=1}^\infty(-1)^s\exp\left[-2\left(\frac{\pi\Gamma s}{\hbar\omega_c}\right)^l\right]\frac{s2\pi^2k_BT_e/\hbar\omega_c}{\sinh(s2\pi^2k_BT_e/\hbar\omega_c)}\nonumber\\
    &&\times\cos\left(\frac{2\pi E_Fs}{\hbar\omega_c}\right)\cos\left(\frac{\pi s\Delta_s}{\hbar\omega_c}\right).
\label{equSdH}
\end{eqnarray}

Eq.\ref{equSdH} comprises the case of dominant long-range ($p=1$)
and short-range ($p=2$) scattering mechanism, Lorentzian ($l=1$)
and Gaussian ($l=2$) LL broadening. We note that the LL shape is
not unambiguously determined from Eq.~(\ref{equSdH}): a
$B$-independent Lorenzian broadening $\Gamma_L$ is equivalent to a
$B$-dependent ($\Gamma_G\propto\sqrt{B}$) Gaussian broadening
$\Gamma_G$, where $\Gamma_L=\frac{\pi}{\hbar\omega_c}\Gamma_G^2$.
The terms ``Lorentzian'' and ``Gaussian'' broadening are used here
in the sense of magnetic field independent broadening.

Finally, an arbitrarily large spin splitting $\Delta_s$ is taken
into account by the last cosine term.
\cite{Fang68,StephensPRB11-4999-1975,KurganovaPRB84-121407-2011}
In CdTe, at low magnetic fields, the spin splitting is much
smaller than the cyclotron energy such that $\cos\left(\frac{\pi
s\Delta_s}{\hbar\omega_c}\right)\approx 1$, which does not
influence much the SdH amplitude. In contrast, in CdMnTe,
$\Delta_s$ at low magnetic field can be much larger than
$\hbar\omega_c$, which leads to the beating patterns described in
the previous section. In this case, the cosine term in
Eq.~(\ref{equSdH}) describes the additional modulation of the
envelope of the SdH oscillations.

\section{Data modeling}

\subsection{SdH in CdTe}

\begin{figure}[h]
\centering
\includegraphics[width=8.5cm]{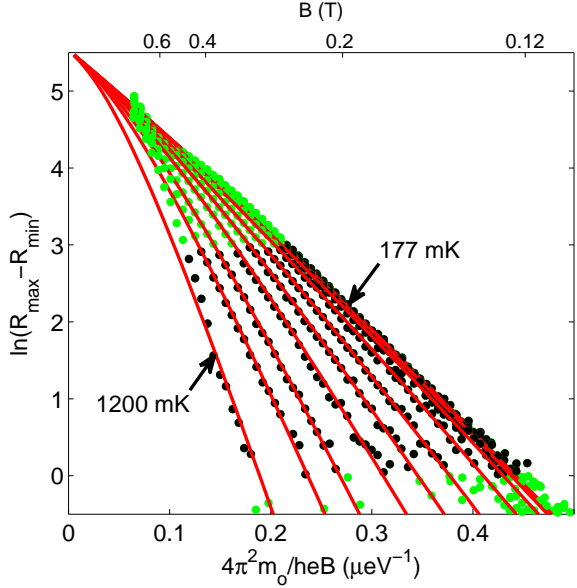}
\caption{Natural logarithm of the amplitude of SdH oscillations in
CdTe for several carrier temperatures from 177 to 1200~mK (full
circles). Due to the limited range of validity of Eq.~(\ref{equSdH})
and/or the smallness of signal-to-noise ratio, only the black
points were used in the fitting procedure. The red curves are the
theoretical fit using Eq.~(\ref{equSdH}) with a Landau level
broadening $\Gamma=110~\mu$eV and an effective mass $m_e=0.1m_0$.}
 \label{Fig2}
\end{figure}

The amplitude of the SdH oscillations in CdTe is plotted in
Fig.~\ref{Fig2} for carrier temperatures from 177~mK to 1200~mK.
The amplitudes are compared with the model Eq.~(\ref{equSdH}),
 using a Landau level broadening $\Gamma=112\pm10~\mu$eV, $m_e=0.1m_0$, $R_0=65\Omega$, and $g_e=-1.6$.
The carrier temperature $T_e$ which differs from the bath
temperature in our experimental conditions was determined as a
fitting parameter and is used throughout the paper. The results
were found to be essentially the same when taking into account one
or more terms in the Fourier series of Eq.~(\ref{equSdH}). The data
are well-described with a long-range scattering formalism ($p=1$)
and Lorentian Landau levels ($l=1$). The dominant role of
long-range scattering mechanism, as well as the extracted value
for the quantum life time $\tau_q=(3\pm0.3$)~ps is a fingerprint
of a good sample quality, sufficient to observe the integer and
fractional quantum Hall effects in this II-VI semiconductor
material as reported in our earlier study. \cite{Piot10}

\subsection{SdH in CdMnTe}

The SdH oscillations in CdMnTe are analyzed in two steps. We first
focus on the position of the nodes (section \ref{nodes}), and then
discuss the oscillation amplitude (section \ref{ampCdMnTe}). The
final overall behaviour is summarized in section
\ref{finalsumulation}.

\subsubsection{Characteristics of the SdH ``nodes''}\label{nodes}

\begin{figure}[h]
\centering
\includegraphics[width=9cm]{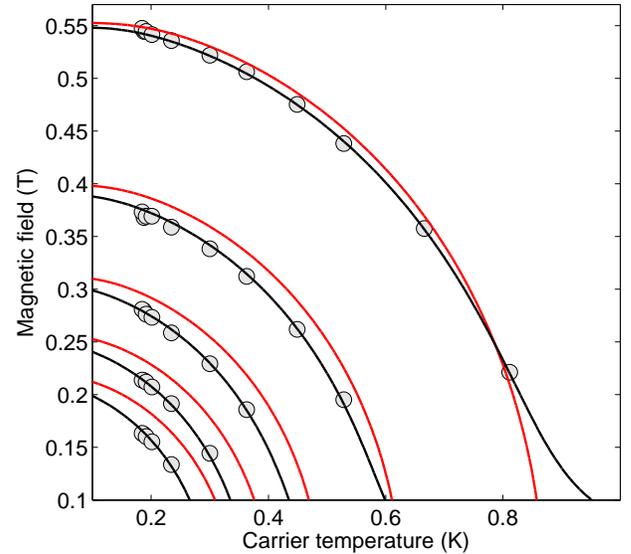}
\caption{Magnetic field position of the nodes in the beating
pattern of $R_{xx}$ in CdMnTe QW plotted versus carrier
temperature. Experimental data (points) are compared with the
models of GZS including/neglecting Mn pair clusters (black/red
curves). The mean AF exchange interaction was included by the
parameter $T_{0}=(40\pm10$)~mK and $T_{0}=(90\pm10$)~mK in the two
models of GZS, respectively.}\label{Fig3}
\end{figure}

The temperature dependence of the magnetic-field position of the
SdH nodes is shown in Fig.~\ref{Fig3}.
As explained in section \ref{expRes}, the nodes appear when the
condition $\Delta_s=(n+\frac{1}{2})\hbar\omega_c$ is fulfilled.
The spin splitting $\Delta_s$ in CdMnTe can be modelled in a
mean-field approach as:
\begin{eqnarray}
\lefteqn{\Delta_s=g_e\mu_BB+\Delta_{exch}\mathcal{B}_{5/2}\left[\frac{\frac{5}{2}g_{Mn}\mu_BB}{k_B(T_{Mn}+T_{0})}\right](1-P_p)+}\nonumber\\
&&\frac{\Delta_{exch}P_p}{2S_0}\sum_{n=1}^5\frac{1}{\exp\left(\frac{2nJ_{AF}-g_{Mn}\mu_BB}{k_BT_{Mn}}\right)+1}+\alpha\Delta_s,
\label{equSpinSplitting}
\end{eqnarray}
where the four terms are the bare Zeeman splitting, the giant
Zeeman splitting (GZS) due to the $s-d$ exchange interaction
between electrons and isolated Mn spins, \cite{Gaj79} the
contribution of anti-ferromagnetic (AF) interactions within pair
clusters of Mn atoms \cite{AggarwalPRB32-1985} which modifies the
average Mn spin polarization, and the contribution of
electron-electron interactions, \cite{Kunc10} respectively. The
strength of the $s-d$ electron-manganese interaction
$\Delta_{exch}$ depends in particular on the Mn concentration
(nominally, $x_{ave}=0.3\%$) and the total Mn spin quantum number
$S_0=5/2$. $\mathcal{B}_{5/2}$ is the Brillouin fonction
describing the Mn spin magnetism, where $g_{Mn}=2.0$ is the
manganese $g$-factor. $T_0$ is an additional phenomenological
temperature which can be introduced to take into account AF Mn-Mn
interactions through the (single particle) Brillouin function.
$T_{Mn}$ is Mn temperature (we note that the best fit to the data
were obtained with $T_{Mn}=T_e$). The ``cluster'' term aims at
direcly including the effect of AF Mn-Mn interactions on the spin
polarization of the $S_0=5/2$ Mn system. $P_p$ is the probability
that Mn is a part of a pair cluster, and $J_{AF}$ is the strength
of the direct Mn-Mn AF interaction between two neighbors.
$\alpha\cong\frac{\Delta_0}{2E_F}=0.11$ stands for the
electron-electron interaction (in our case, the Fermi energy is
$E_F=9.6$~meV and the parameter $\Delta_0=2.1$~meV has been
determined in previous work. \cite{Kunc10})

We have used two different approach to fit the experimental data
of Fig.\ref{Fig3}. In the first approach, the e-Mn interaction is
entirely taken into by the second term of
Eq.\ref{equSpinSplitting} and the third term is neglected
($P_p=0$). The best fit to the experimental data are reported as
red curves in Fig.\ref{Fig3}, and describes the data at the
qualitative level. The corresponding parameters are $\Delta_{exch-1}=1.7$~meV and
$T_{0-1}=90$~mK, which phenomenologically takes into account the
Mn-Mn interaction by reducing the average Mn spin
polarization.

In the second approach, the Mn-Mn AF interaction within pair
clusters is directly taken into account by using the third
(``cluster'') term of Eq.\ref{equSpinSplitting}. In this term, the
nearest-neighbor (NN) AF interaction is expected to be rather
strong ($J_{NN}/k_B\approx5$K \cite{Gaj79,AggarwalPRB32-1985}),
such that for magnetic fields lower than
5~T(=$(2J_{NN}-5k_BT_{e,max})/(g_{Mn}\mu_B)$, where
$T_{e,max}=823$~mK), NN are always anti-ferromagnetically coupled
and thus do not contribute to the GZS. However, the next nearest
neighbors (NNN) of manganese atoms interaction is weaker
($J_{NNN}/k_B=0.5$K,\cite{KreitmanPhysRev144-367-1966,BastardPRB20-4256-1979,ShapiraPRB30-1984,NovakJAP57-3418-1985})
and can play a role already at $B\sim0.5$T and $ T\sim 200$mK and
similar or lower $B/T$ values. We note that the interaction
strength between third and higher order NN is generally small and
decreases exponentially with distance
.\cite{NovakJAP57-3418-1985,LarsonPRB37-1988} Their residual
influence is sufficiently well-described by the commonly used
$T_0$  phenomenological parameter. \cite{Gaj79} Besides these
distant pairs, AF interactions from higher order clusters
(triplets, quadruplets, etc.) are also included in $T_0$. The fit
to the data obtained in this second approach is plotted as black
curves in Fig.~\ref{Fig3}, and gives an excellent quantitative
description of the data. The fitting parameters are $P_p=20\%$,
$J_{AF}/k_B=0.5$K, $\Delta_{exch-2}=1.7$~meV and
 $T_{0-2}=40\pm10$mK. The extracted probability of clusters
formation, $P_p=20\%$, is significantly higher than the one
expected from statistical considerations (typically a few
percent). This is usually explained in terms of non-homogeneous
distribution of Mn.\cite{GalazkaPRB22-1980} The value of
$J_{AF}/k_B=0.5$K suggests that the influence of the Mn pair
clusters originates from the next nearest neighbors (NNN)
interaction of manganese ions. The small but non-zero value
observed for $T_{0-2}=40\pm10$mK can be attributed to higher order
AF interactions.

As a conclusion, the beating pattern of the SdH oscillations is
profoundly modified by the magnetic sub-system, and therefore
constitutes a powerful tool to characterize the e-Mn interaction,
the Mn concentration, as well as the Mn-Mn interactions in CdMnTe
systems.

\subsubsection{Oscillation amplitude}\label{ampCdMnTe}

\begin{figure}[h]
\centering
\includegraphics[width=8.5cm]{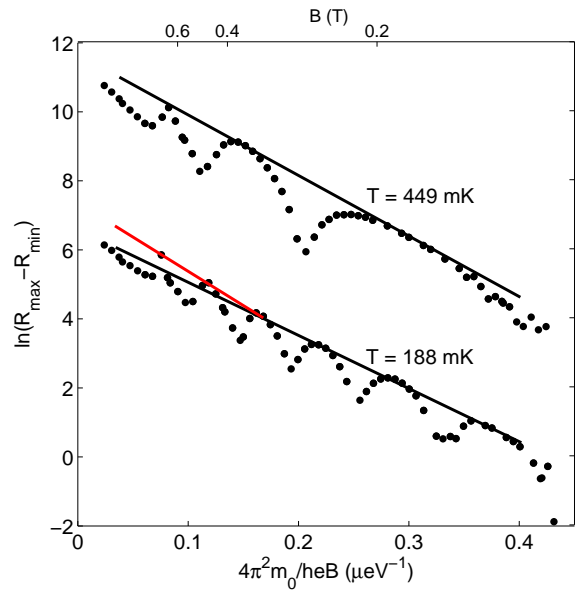}
\caption{Natural logarithm of SdH oscillations amplitude in CdMnTe
at the electron temperatures $188$ and $449$~mK, in reciprocal
magnetic field scale. The two straight black lines shows the usual
linear $1/B$ behavior expected within the Lifshitz-Kosevich
formalism, nevertheless exhibiting an unusual parallelism for
different temperatures. The red line depicts nonlinearity,
described in our model as a field dependent Landau level
broadening. The data are shifted vertically for clarity.
}\label{Fig4}
\end{figure}

In Fig.~\ref{Fig4}, we plot the SdH oscillation amplitude in our
CdMnTe QW for two representative temperatures, using a reciprocal
magnetic field scale.
In addition to the previously analyzed beating patterns, which
manifest themselves as repeated deviations from the solid black
line, we observe that the overall envelope of the amplitude of the
oscillations depends only weakly on temperature. This is
particularly evident when comparing the almost parallel solid
black lines in Fig.~\ref{Fig4} to the case of CdTe
(Fig.~\ref{Fig2}) where the $1/B$ slope of the oscillation
amplitude is strongly increasing with temperature, a usual
consequence of the SdH temperature damping. This suggests that the
usual SdH temperature damping, described by the `` $x/sin(x)$''
function in Eq.~(\ref{equSdH}), is compensated by some non-trivial
temperature dependence of the disorder damping (exponential term
in Eq.~(\ref{equSdH}). More precisely, the observed behavior points
toward an unusual temperature narrowing of the LL broadening (at
thus a reduced ``disorder damping'' at higher temperature) which
cannot be anticipated within the most standard forms of
scattering. Another interesting observation is the non-linearity
of the overall envelope in the $1/B$ scale, which is indicated at
low temperature by the red line in Fig.~\ref{Fig4}. This $1/B$
non-linearity implies a magnetic field dependence of the Landau
level broadening $\Gamma$.

We note that the value of quantum lifetimes in CdTe and CdMnTe are
here very similar, in agreement with the previous observations.
\cite{Piot10,CdMnTeFQH2014} However, the observed field and
temperature dependence of this quantity in CdMnTe points out to
the occurrence of additional physical effect contributing to the
level broadening. Again, the main difference between CdTe and
CdMnTe QWs is the presence of manganese spins. As a matter of
fact, the observed dependence of the broadening are qualitatively
reminiscent of the behavior of the average manganese spin
polarization $\langle S_z\rangle$, which increase (decreases) with
the magnetic field (temperature), as can be seen by considering
the (dominant) Brillouin function in Eq.~(\ref{equSpinSplitting}). As
the spin polarization of the manganese system directly determines
the position of the Landau level, a \textit{variation} of the
manganese content $x$ which appears in the pre-factor
$\Delta_{exch}$ in Eq.~(\ref{equSpinSplitting}) will shift the energy
level position proportionally to $\langle S_z\rangle$. A
non-homogenous Mn distribution at the local scale is therefore at
the origin of an additional level broadening. The mean energy
shift $\Gamma_{Mn}$ can be simply written from
Eq.\ref{equSpinSplitting} as:

\begin{eqnarray}
\Gamma_{Mn}= \left( \frac{\Delta
x}{x_{ave}}\right)\left(\frac{1}{2}\Delta_{exch}\right)
\mathcal{B}_{5/2}\left[\frac{\frac{5}{2}g_{Mn}\mu_BB}{k_B(T_{Mn}+T_0)}\right],
\label{equSpinSplitting}
\end{eqnarray}

where $\Delta x$ represent the maximum Mn spatial fluctuation
around the average value $x_{ave}$ (the extremal values of $x$ are
then $x=x_{ave} \pm \Delta x$). The other parameters are the ones defined previously, and the mean energy shift $\Gamma_{Mn}$ can be identified with the Lorentzian
HWHM used in our formalism Eqs.~(\ref{lor}) and (\ref{equSdH}).

\begin{figure}[h]
\centering
\includegraphics[width=8.5cm]{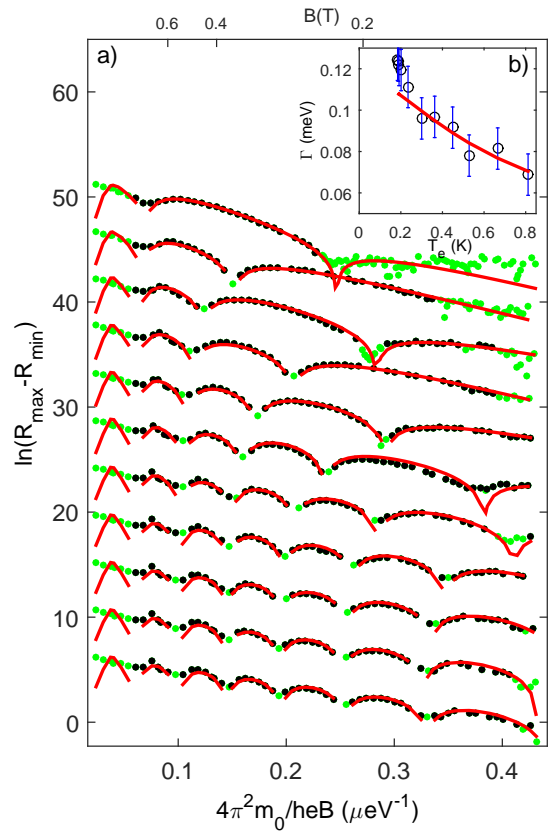}
\caption{Natural logarithm of the amplitude of SdH oscillations in
CdMnTe QW for electron temperatures from 188 to 823~mK.
Experimental data used (unused) in the fitting procedure are
plotted by black (green) points and the red curves show the
theoretical model including fluctuations of the manganese
concentration as a additional source of Landau level broadening.
(inset) Resulting total LL broadening $\Gamma$ as a function of
the magnetic field $B$ and the carrier temperature~$T$.}
\label{Fig5}
\end{figure}

We have reproduced our experimental data by taking this
effect into account and writing the total LL in CdMnTe broadening
as: $\Gamma=\Gamma_0+\Gamma_{Mn}$, where $\Gamma_0$ is a
temperature/field independent broadening, and $\Gamma_{Mn}$ is the
``fluctuation-induced'' contribution described above. This
broadening was directly injected in the previously used SdH
formalism (Eq.\ref{equSdH}). The amplitude of SdH oscillations is shown in
Fig.~\ref{Fig5} together with simulations using a LL broadening
$\Gamma=\Gamma_0+\Gamma_{Mn}$ (red solid curves).
As in the case of CdTe, we have used Lorentzian Landau levels in a
long-range scattering approximation ($p=1$ in Eq.\ref{equSdH}).
The data were fitted using $\Gamma_0=20~\mu$eV and
$\frac{\delta x}{x_{ave}}=(11\pm5)$\% which
gives an estimation for the relative mean fluctuation of the Mn
concentration. This value is in good agreement with expected statistical fluctuation of number of manganese ions $N_{Mn/e}$ per one electron (or per area defined by de Broglie wavelength). In our sample $N_{Mn/e}\approx60$, giving fluctuations $\frac{1}{\sqrt{N_{Mn/e}}}=13\%$.

The resulting total LL broadening $\Gamma$ is plotted as a
function of magnetic field and temperature in the inset of figure
\ref{Fig5}. The extracted ``non-magnetic'' broadening $\Gamma_0$
is smaller in CdMnTe than in CdTe. This is actually not
surprising, because our approach considers an \textit{additional}
source of broadening in CdMnTe while the total broadening are
similar in both systems. Physically, a smaller $\Gamma_0$
broadening in CdMnTe could be attributed to a reduction of the
intra-Landau level spin-flip scattering in CdMnTe. Indeed, while
the opposite spin levels in CdTe always belong to the same Landau
level, in CdMnTe the GZS puts into coincidence opposite spins with
different orbital quantum number which might affect the spin-flip
scattering processes. 

We finally note that the value of $\frac{\delta
x}{x_{ave}}=(11\pm5)$\% is obtained by assuming that
the temperature/field dependent Landau level broadening originates
\textit{only} from fluctuation in the manganese concentration, and
thus constitutes an upper bound for the mean fluctuations. Other
mechanisms involving the Manganese spin polarization $\langle
S_z\rangle$, such as an anisotropic electron-Mn interaction
similar to the Mn-Mn anisotropic Dzyaloshinski-Moryia interaction,
\cite{SamarthPRB37-9227-1988} could also contribute to the
observed broadening.

\subsubsection{Overall behaviour}\label{finalsumulation}

\begin{figure}[h]
\centering
\includegraphics[width=8.5cm]{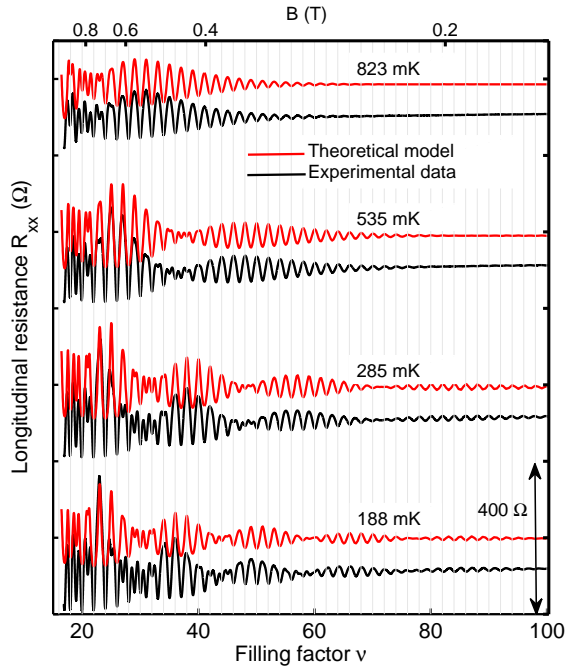}
\caption{Longitudinal resistance $R_{xx}$ in the CdMnTe QW (black
curves) compared with our model (red curves) assuming manganese
pair clusters and fluctuations of the manganese
concentration.}
\label{Fig6}
\end{figure}

In Fig.~\ref{Fig6}, we report the experimental data of Fig1.b,
together with the model developed throughout the paper for the
temperature and field dependence of the SdH oscillation in CdMnTe.

The oscillations amplitude and the nodes position are well
reproduced as shown earlier in the paper. The beating pattern
together with the phase shift of the oscillations across each node
are correctly described by the last cosine term in
Eq.~(\ref{equSdH}) including the giant Zeeman splitting. We note
that in order to reproduce the doubled oscillation frequency at
the high-field nodes ($\nu<50$), the first 2 terms of the Fourier
series in Eq.~(\ref{equSdH}) had to be taken into account. Taking
into account higher order terms (up to 100) deepen the splitting
of the doubled SdH oscillations, which is nevertheless still
weaker than in our experimental observations. This could be
related to the proximity of the Stoner transition where spin
splitting develops in a non-linear way.
\cite{PiotPRB72-2005,PiotPRB2007}

\section{Conclusions}

Shubnikov De-Haas oscillations have been studied in a high quality
magnetic 2DEG formed in a diluted magnetic CdMnTe quantum well.
The SdH characteristics can be well-described by incorporating the
electron-Mn exchange interaction into the traditional
Lifshitz-Kosevich formalism in a ``mean field'' approach. A more
detailed analysis reveals the role of antiferromagnetic Mn-Mn
interactions in this system, as well as a non trivial reduction of
the Landau level broadening with increasing temperature which
could be accounted for by fluctuations in the manganese
concentration.

We thank P. Kossacki, M. Goryca, and M. Orlita for helpful
discussions. This work was part of the research program
MSM0021620834 financed by the Ministry of Education of the Czech
Republic. The research in Poland was partially supported by National Science Centre (Poland) grant DEC-2012/06/A/ST3/00247.

\end{document}